\documentclass{ws-procs961x669}
\usepackage{graphicx}
\usepackage{chapterbib}
\def\be{\begin{equation}}
\def\ee{\end{equation}}
\def\bea{\begin{eqnarray}}
\def\eea{\end{eqnarray}}
\begin{document}

\title{Event horizons under the effect of the Penrose process}

\author{Hernando Quevedo}

\address{Instituto de Ciencias Nucleares, Universidad Nacional Aut\'onoma de M\'exico, \\ AP 70543, Mexico City, Mexico.\\
Dipartimento di Fisica and Icra, Universit\`a di Roma “La Sapienza”, Roma, Italy. \\
Al-Farabi Kazakh National University, Al-Farabi av. 71, 050040 Almaty, Kazakhstan.}

\date{\today}

\begin{abstract}
We investigate how test particles absorbed by a black hole affect the properties of the event horizon. We consider particles that arrive from infinity with positive energy and cross the horizon. We also study the absorption of particles with negative energy, which are generated inside the ergosphere as the result of the decay of other particles, following the dynamics of the Penrose process. We show that, in general,  the absorption process leads to an increase in the horizon radius and, consequently, of any physical quantity that is proportional to the horizon radius, such as the irreducible mass or the entropy. 

\end{abstract}

\keywords{Particle absorption, event horizons, Penrose process}

\bodymatter

\section{Introduction}
\label{sec:int}

One of the most interesting results of Einstein's theory of gravity is the prediction of the existence of black holes, a prediction that recently has been observationally confirmed. In fact, in 2019, 
 the first direct image of the vicinity of a black hole was published, following observations made by the Event Horizon Telescope in 2017 of the black hole in Messier 87's galactic center \cite{event2019first}. This result has increased the interest in studying the physical properties of black holes not only from a theoretical but also from an observational viewpoint.

A particular process that has been analyzed from several points of view is the fall of a test particle in the gravitational field of a black hole \cite{thorne2000gravitation}. In this case, particular attention has been paid to investigating the radiation emitted by a test particle falling radially into a non-rotating black hole \cite{davis1971gravitational}. In the case of a rotating black hole, the gravitational waves emitted by a falling spinning massive particle have been analyzed by using standard approximate methods 
\cite{mino1996gravitational}.

In this work, we analyze the case of a test particle that falls without emission of radiation. Instead, we assume that the particle is completely absorbed by the black hole, leading to a change in its state.  
It is expected that the mass and angular momentum of the in-falling particle would affect also the corresponding characteristics of the black hole, namely, its mass and angular momentum as well as any other quantities that depend on these two parameters. 

Since the mass and angular momentum of the test particles are small quantities, we will assume that the change of the black hole parameters depends linearly on the parameters of the test particle. This implies that during the absorption of a test particle no exchange  occurs between the rest energy and rotational energy of the black hole. 

We are interested in analyzing how the absorption of a test particle affects the properties of the black hole event horizon, which, in turn, depends on the mass and angular momentum parameters of the black hole. 

We will consider first test particles that arrive from infinity with positive energy and cross the event horizon. We will also analyze the case of particles with negative energy (Penrose particles) that are generated inside the ergosphere using the Penrose process and then enter the horizon, modifying the properties of the black hole. 

This work is organized as follows. In Sec. \ref{sec:bhs}, we review the main properties of the Kerr spacetime, emphasizing the role of the event horizons and the ergosphere. Then, in Sec. \ref{sec:abs}, we describe the process of absorption of a test particle by a black hole, considering explicitly the influence of this process on the outer event horizon. 
We consider static and stationary black holes and show that the absorption process, in general, leads to an increase in the outer horizon radius. 
In Sec. \ref{sec:many}, we analyze the process of many-particle absorption, in which particles are released one by one into the black hole,  taking into account the condition that the sequence of test particles does not lead to the formation of naked singularities. 
Finally, in Sec. \ref{sec:con}, we summarize our results and formulate their consequences for analyzing the behavior of quantities that depend explicitly on the horizon radius.


\section{The Kerr spacetime}
\label{sec:bhs}

Consider the spacetime of a rotating black hole described by the Kerr metric, which in Boyer--Lindquist coordinates and geometrized units $G = 1 = c$ can be written as 
\bea\nonumber
 ds^2=-\frac{(\Delta -a^2\sin^2\theta)}{\Sigma
}dt^2-\frac{2a\sin^2\theta (r^2+a^2-\Delta)}{\Sigma}dtd\phi+
\eea
\be\label{knnetric}+\left[\frac{(r^2+a^2)^2-\Delta
a^2\sin^2\theta}{\Sigma}\right]\sin^2\theta
d\phi^2+\frac{\Sigma}{\Delta}dr^2+\Sigma d\theta^2, \ee
where
\be
\Delta = r^2-2M r+a^2,
\ee
and
\be
\Sigma =r^2+a^2\cos^2\theta.
\ee
The parameters $M$ and $a=J/M$ stand for the gravitational mass and angular momentum $J$ per unit mass  of the gravitational source, respectively, as measured
by a distant observer.

We can see that the metric is ill-defined at those locations where the conditions  $\Sigma = 0$ and $\Delta =0$ are satisfied. The first case $\Sigma = r^2+a^2\cos^2\theta =0$ can be satisfied only for $\theta = \pi/2$ and $r=0$, which is interpreted as a ring-like curvature singularity located on the equatorial plane. The second condition, $\Delta = r^2-2M r+a^2 =0$, can be shown to correspond to a coordinate singularity, which determines the event horizons located at the radii
\be
R_{H\pm} = M \pm \frac{\sqrt{M^4 - J^ 2 } }{M} .
\label{rh}
\ee
under the condition that $M^4\geq J^2$. If this condition is not satisfied, no horizon exists, and the above metric describes the gravitational field of a naked singularity.

\begin{figure}
    \centering
    \includegraphics[scale=0.8]{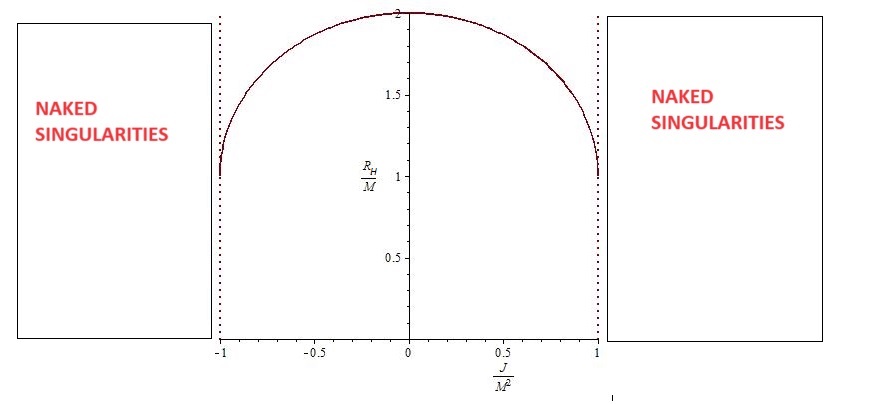}
    \caption{Behavior of the outer horizon radius $R=R_{H+}$ as a function of the angular momentum $J$ of the Kerr black hole. The maximum value $R=R_S=2M$ corresponds to the static limiting case of a Schwarzschild black hole ($J=0$), whereas the minimum value is reached for the extreme Kerr black hole with $J=M^2$. If the black hole condition $J\leq M^2$ is not satisfied, naked singularities are formed. }
    \label{fig1}
\end{figure}

In Fig. \ref{fig1}, we illustrate the behavior of the outer event horizon radius 
\be 
R_H=R_{H+}=M + \frac{\sqrt{M^4 - J^ 2 } }{M} 
\label{rhp}
\ee
for different values of the angular momentum. We see that the radius is defined within the interval  $M\leq R_H\leq 2M$,  whose limits correspond to the extreme Kerr black hole  $R_{extreme}=M$ and the static Schwarzschild black hole $R_S=2M$, respectively.

Another important surface called ergosphere is determined by  the condition $g_{tt}=0$, whose solutions define the ergosphere radii \cite{1971ESRSP..52...45R} 
\be
R_{E\pm} =  M \pm \frac{\sqrt{M^4 - J^ 2 \cos^2\theta} }{M} .
\label{re}
\ee

The  limiting cases contained in the Kerr spacetime are the Schwarzschild metric, which is recovered for $a=0$, and
the Minkowski metric of special relativity for $a=M=0$, a result that reinforces the physical significance of the Kerr solution.  

An important property of the ergosphere region is that the effective potential that governs the motion of test particles allows the possibility of states with negative energy \cite{1971ESRSP..52...45R}, which is an important condition for the realization of the Penrose process \cite{1969NCimR...1..252P}. Indeed, if we consider a test particle falling from far away into a Kerr black hole, once it enters the ergosphere, it can split into two particles, one with positive energy and the second one with negative energy. The particle with negative energy is then absorbed by the black hole, leading to a change in its state, and the particle with positive energy returns to infinity with an energy that is larger than the energy of the incoming particle. In this way, energy can be extracted from the black hole, which is the main aspect of the Penrose process.  

\section{Absorption of test particles}
\label{sec:abs}

Consider a test particle of mass $m$ and angular momentum $j$ falling into the gravitational field of a rotating black hole of mass $M$ and angular momentum $J$. The test particle condition implies that $m<<M$ and $j<< J$. The absorption of a test particle implies a change in the state of the black hole, which undergoes a transition from an initial state characterized by the parameters $M_i$ and $J_i$ to a final state with parameters $M_f$ and $J_f$. 

Due to the smallness of the parameters $m$ and $j$ of the test particle, we can assume that the final and initial states of the black hole are connected by the simple relationship 
\be 
M_f = M_i + m , \quad J_f = J_i + j.
\ee

Any quantity that depends on $M$ and $J$ will be affected by the absorption of a test particle. Consider, for instance, the outer horizon radius $R_{H}$. The initial and final states are denoted by $R_i$ and $R_f$, respectively. We introduce the change $\Delta R$ of the horizon radius as 
\bea
\Delta R_H & = & R_{H,f} - R_{H,i} \nonumber\\
& = &  
m + \frac{1}{M_i + m} \sqrt{(M_i+m)^4 - (J_i+j)^2}  -\frac{1}{M_i}\sqrt{M_i^4 - J_i^2} .
\label{Delta}
\eea 
For this quantity to be physically meaningful,  we must assume that the initial state is that of a black hole, i.e., $M_i^2 \geq J_i$. The final state can be either a black hole or a naked singularity, depending on the initial state and the values of $m$ and $j$. We will explore the case of a final state corresponding to  black holes, i.e., we demand that the condition 
\be
(M_i+m)^4 \geq (J_i + j )^2 
\label{bhc}
\ee
be satisfied. For a given initial state, this condition represents a relationship between the mass and angular momentum of the absorbed particle. 

\subsection{Static black holes}

Consider first the case in which the initial state corresponds to a static black hole with $M_i=M$
and $J_i=0$ and the test particle is spinless. Then, from Eq.(\ref{Delta}), we obtain
\be
\Delta R_H = 2m \ ,
\ee
which is positive for positive values of the test particle mass. This means that the horizon radius is an increasing function of the mass of the incoming test particle. 

\begin{figure}
    \centering
    \includegraphics[scale=0.5]{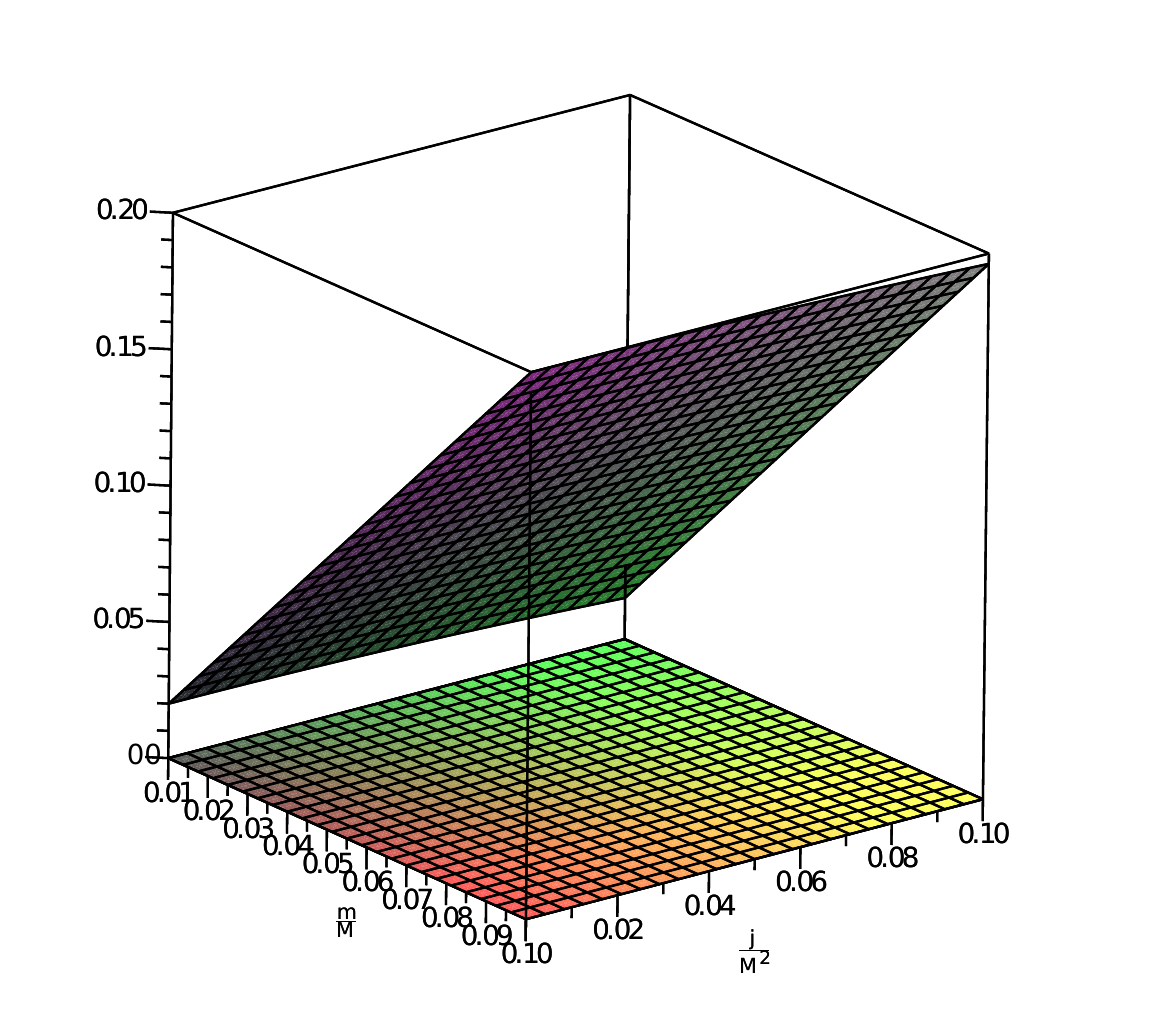}
    \caption{Change of the horizon radius $\Delta R_H$ in terms of the mass $m$ and angular momentum $j$ of the absorbed test particle.}
    \label{fig9}
\end{figure}

Consider now the case of an incoming test particle with mass $m$ and angular momentum $j$. Then, according to Eq.(\ref{Delta}), the change of the horizon radius is given by
\be
\Delta R_H = M-m -  \frac{\sqrt{(M+m)^4-j^2}}{M+m} .
\ee
To evaluate this expression taking into account the test particle character of the angular momentum  $j$, we assume that $j/M^2 << 1$. This assumption also guarantees that the black hole condition $(M+m)^4 \geq j^2$ is satisfied. In
Fig. \ref{fig9}, we illustrate the behavior of $\Delta R_H$ in the interval of allowed values for the parameters of the absorbed particle. It can be seen that $\Delta R_H>0$ in the entire region, implying the horizon radius is an increasing function of the mass and angular momentum of the test particle. 

We conclude that, in general, the absorption of a test particle necessarily leads to an increase in the horizon radius.

\subsection{Stationary black holes}

A test particle that arrives from infinity can have positive mass $m$ with positive or negative angular momentum $j$. However, we can also consider particles that arrive from some point located between the outer horizon and the outer ergosphere, which, according to the dynamics of the Penrose process, allows the possibility of generating particles with negative energy and negative or positive angular momentum. Accordingly, the event horizon of a rotating black hole can be affected by the absorption of test particles with all possible values of $m$ and $j$.

\begin{figure}
\centering
\includegraphics[scale=0.3]{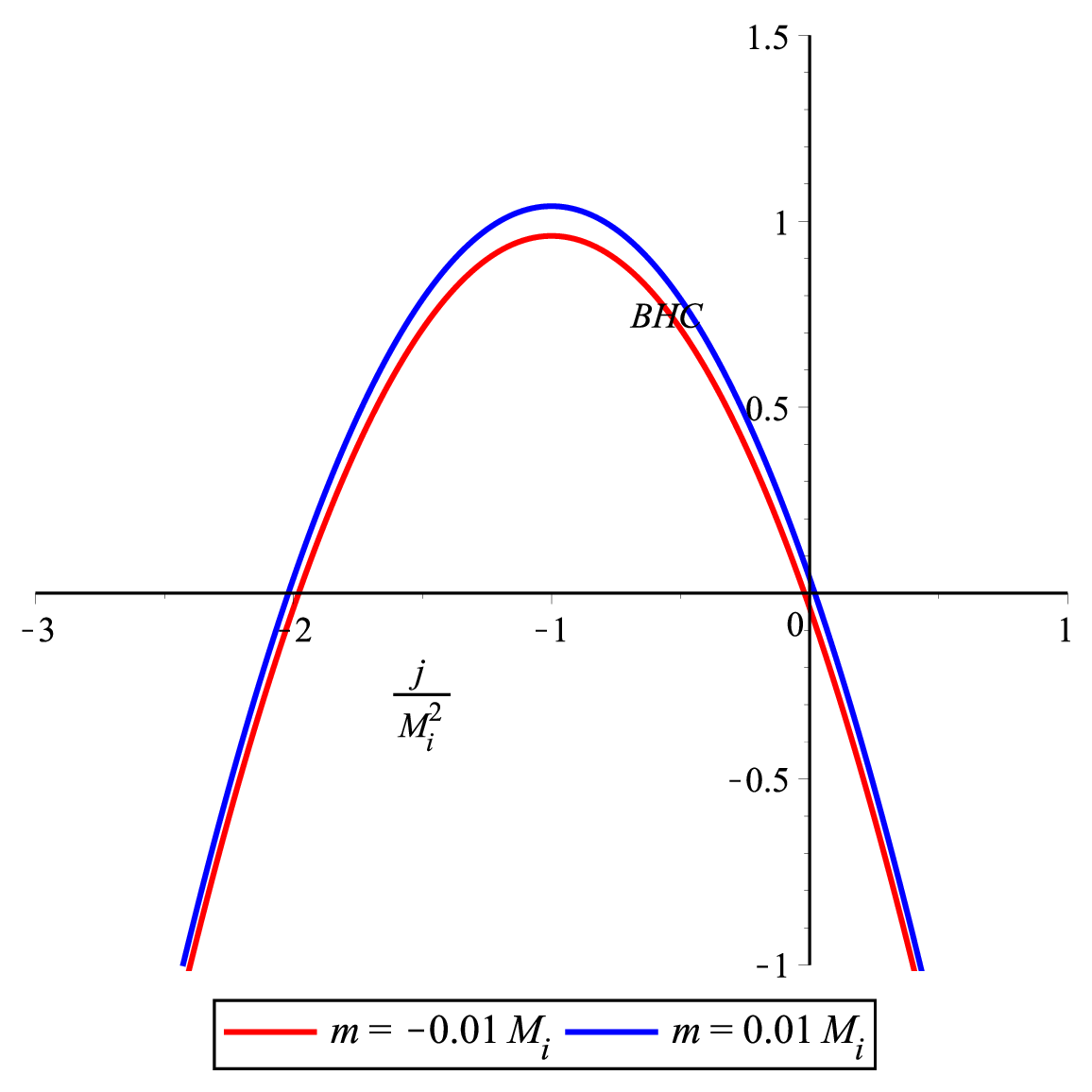}
\includegraphics[scale=0.3]{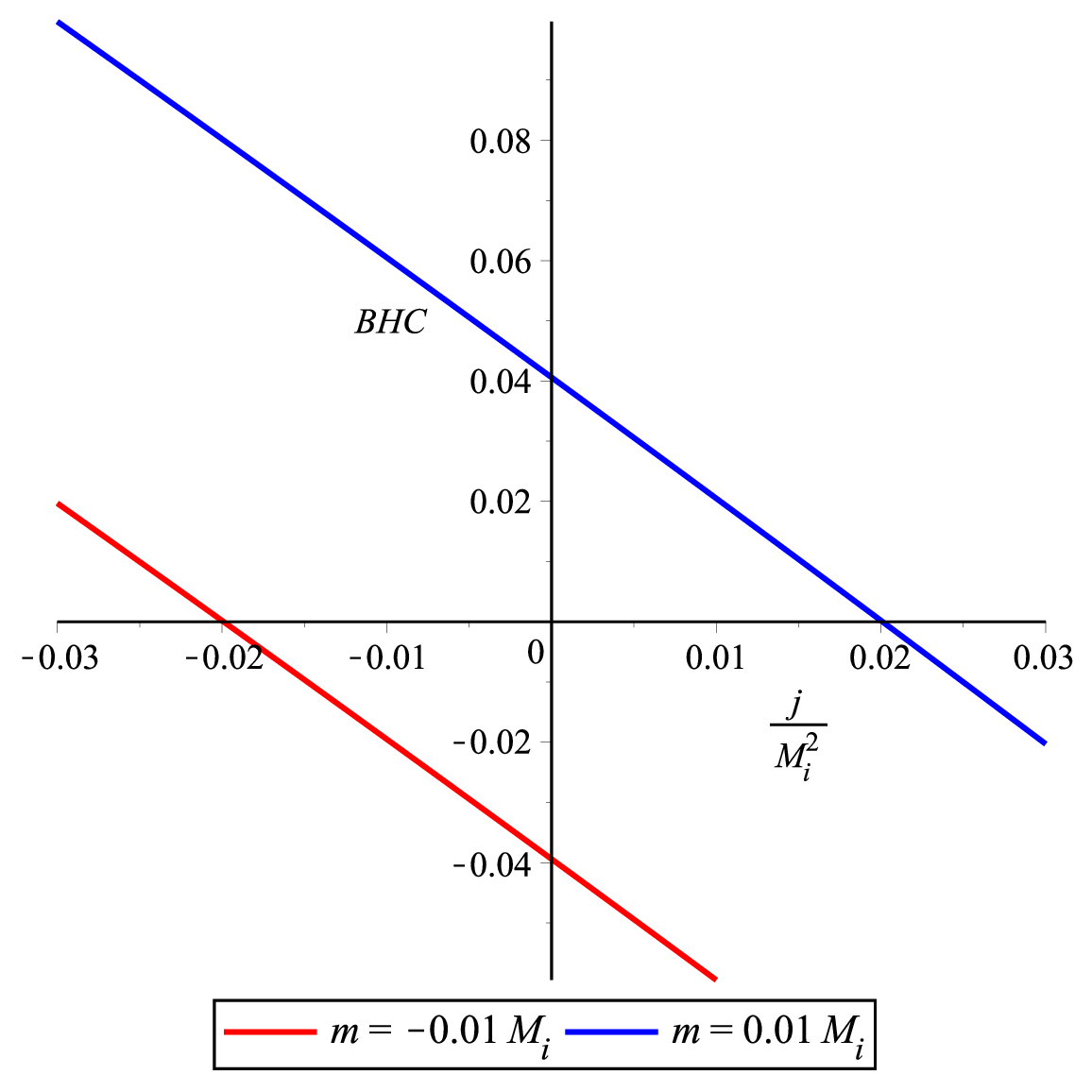}
\caption{The black hole condition for an initial state corresponding to an extreme Kerr black hole.}
\label{fig3}
\end{figure}

Let us consider as initial state an extreme Kerr black hole with $M_i=M$ and $J_i=M^2$. Then, the black hole condition (\ref{bhc}) reduces to 
\be
BHC: \quad 4\,m+6\,{\frac {{m}^{2}}{{\it M}}}+4\,{\frac {{m}^{3}}{{{\it M}}^{2}
}}+{\frac {{m}^{4}}{{{\it M}}^{3}}}-2\,{\frac {j}{{\it M}}}-{\frac {
{j}^{2}}{{{\it M}}^{3}}} \geq 0 .
\label{bhc1}
\ee
In Fig. \ref{fig3}, we plot this condition for fixed values of $m$ and different values of $j$. We see that, in fact, the black hole condition is satisfied only in a small interval of values of $j$. Taking into account the fact that the mass $m$ and angular momentum $j$ of the test particle should be small quantities, we can consider in Eq.(\ref{bhc1}) only linear terms in $m$ and $j$ so that the black hole condition reduces to 
\be
m \gtrsim \frac{j}{2M} ,
\ee
an inequality that should be considered when analyzing the effect of the absorption of test particles.

\begin{figure}
    \centering
    \includegraphics[scale=0.5]{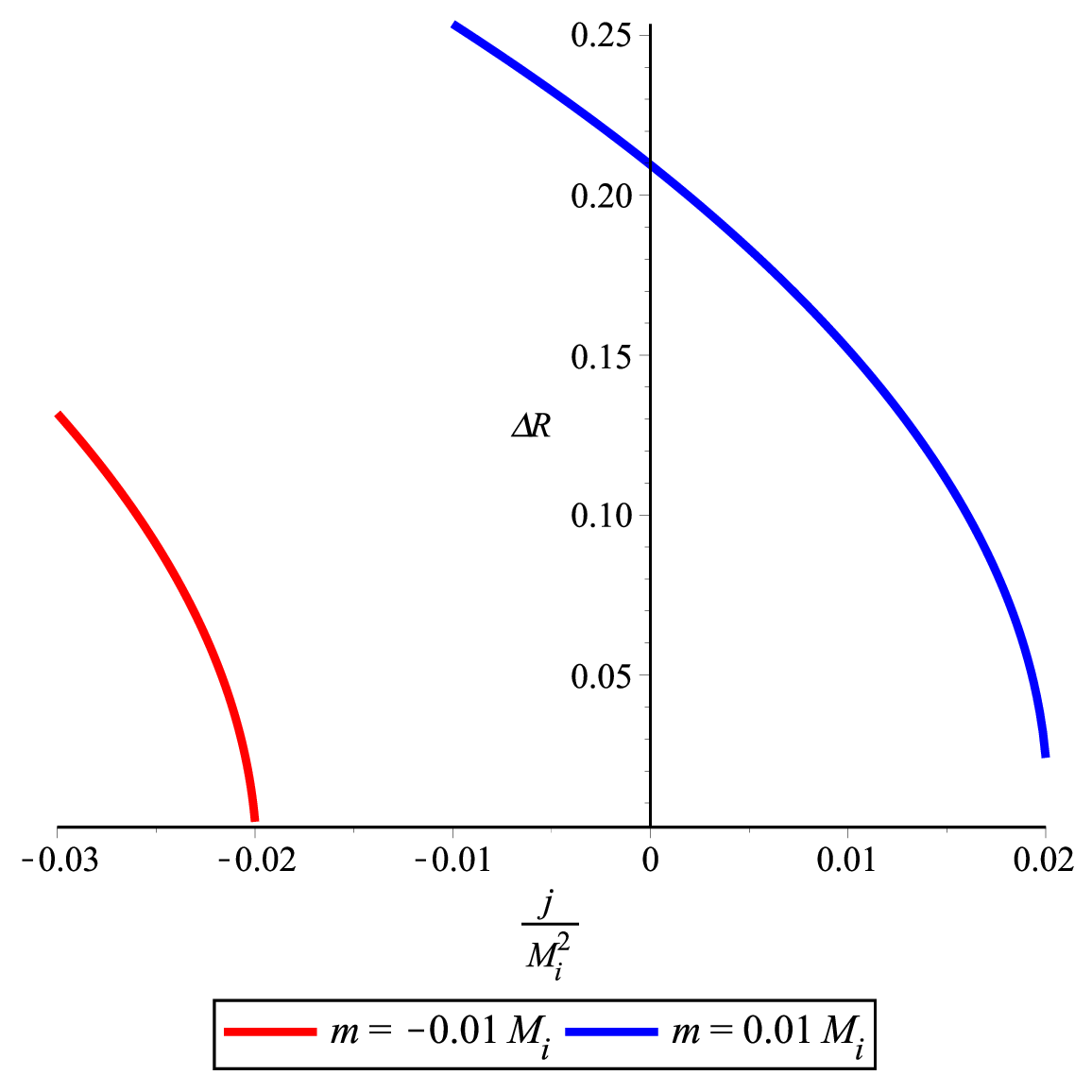}
    \caption{The change of the horizon radius $R_H$ of an extreme black hole under the absorption of a test particle with mass $m$ and angular momentum $j$.}
    \label{fig4}
\end{figure}

We now consider the change in the horizon radius. From Eq.(\ref{Delta}) for an initial extreme Kerr black hole, we obtain
\be
\Delta R_H = m + \frac{1}{M+m}\sqrt{(M+m)^4-(M^2+j)^2} ,
\ee
an expression which is numerically analyzed in Fig. \ref{fig4} for $m=\pm 0.1 M$ in the interval of allowed values of $j$, as demanded by the black hole condition. We see that in the allowed interval the change of the horizon radius is always positive, implying that the absorption process is always accompanied by an increase of the horizon radius.  

\section{Many-particle absorption process}
\label{sec:many}

In this section, we will consider a many-particle process consisting of a sequence of events, in which each event represents the absorption of a particle by the black hole. This process is valid because at each step it is guaranteed that the absorbed particle is a test particle in the sense of the smallness of the physical parameters. 

We will analyze two different types of test particles. First, we study the case of particles that arrive from infinity with positive mass $m$ and positive or negative angular momentum $j$ and cross the event horizon. Secondly, we consider particles that are generated inside the ergosphere following the dynamics of the Penrose process, according to which the particle enters the ergosphere along the equation plane of the Kerr spacetime and then decays into two particles. One of the particles is generated with negative mass $m$ and negative or positive angular momentum $j$ and is then absorbed by the Kerr black hole.

We consider an extreme Kerr black hole as determining the initial state with $M_i=M$ and $J_i=M^2$. Then, in this case, the black hole condition reads
\be
BHC: \quad 4\,nm+6\,{\frac {{(nm)}^{2}}{{\it M}}}+4\,{\frac {{(nm)}^{3}}{{{\it M}}^{2}
}}+{\frac {{(nm)}^{4}}{{{\it M}}^{3}}}-2\,{\frac {nj}{{\it M}}}-{\frac {
{(nj)}^{2}}{{{\it M}}^{3}}} \geq 0,
\ee
where $n$ denotes the number of absorbed particles. The question arises whether this condition can be satisfied for all values of $n$. To answer this question we analyze how this condition depends on the parameters of the test particles. 

\begin{figure}
    \centering
    \includegraphics[scale=0.3]{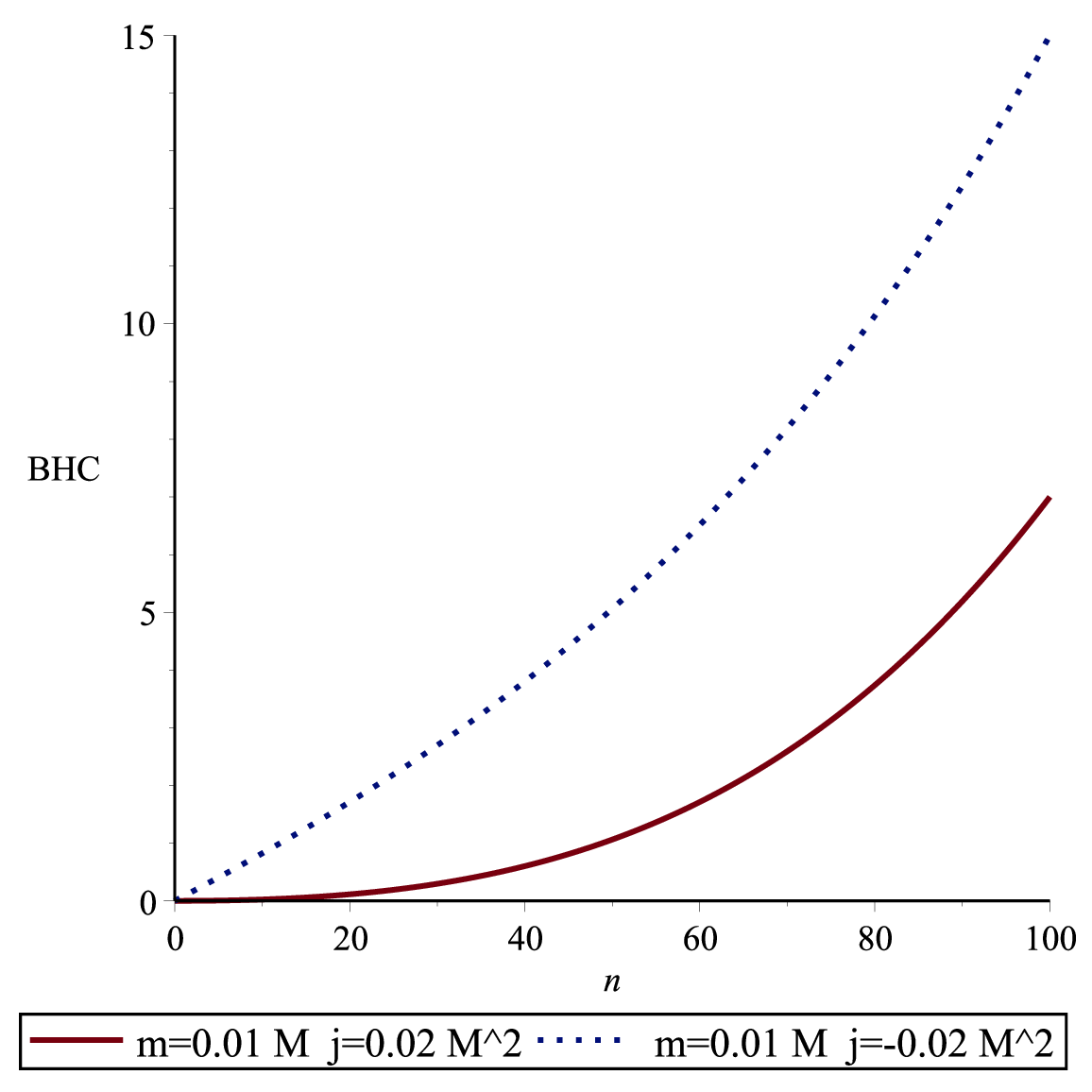}
    \includegraphics[scale=0.3]{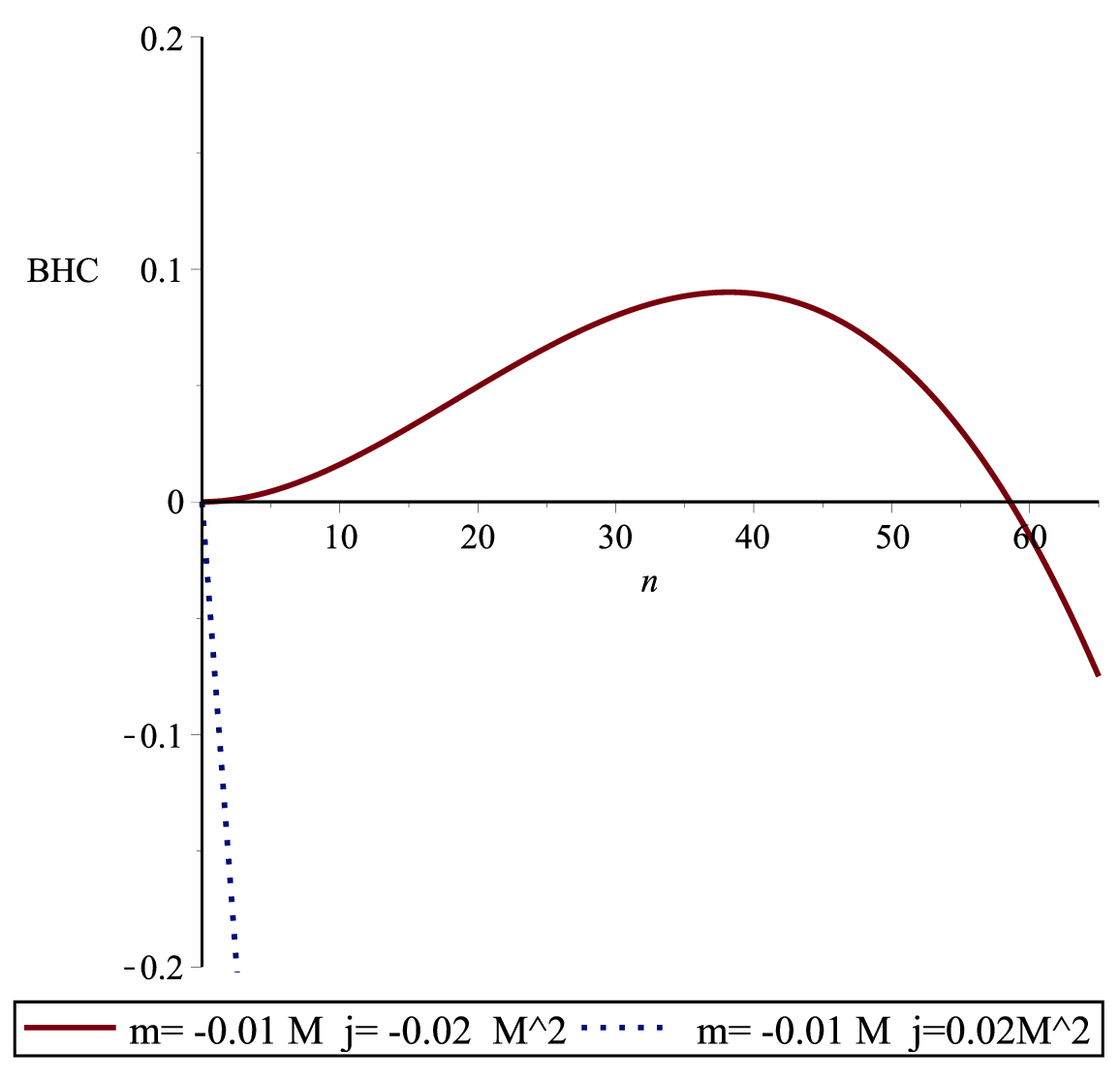}
    \caption{The black hole condition for the absorption of $n$ particles by an extreme Kerr black hole. Left panel: Particles arriving from infinity with mass $m=0.01 M$ and angular momentum $j=\pm 0.02 M^2$. Right panel: Particles arriving from the ergosphere with mass $m=-0.01 M$ and angular momentum $j=\pm 0.02 M^2$.  }
    \label{fig5}
\end{figure}

We illustrate the behavior of the black hole condition in Fig. \ref{fig5}, where the left panel represents particles arriving from infinity with positive mass. In this case, the black hole condition is satisfied for all values of $n$. In the case of Penrose particles coming from inside the ergosphere (right panel) with negative mass, the situation is different. Particles with negative mass and positive angular momentum violate the black hole condition for any values of $n$. This means that this kind of absorption is not allowed. Instead, particles with negative mass and angular momentum can be absorbed without violating the black hole condition, but there is a maximum number of absorptions after which the configuration becomes a naked singularity. In the case shown in Fig. \ref{fig5}, the maximum value is $n=58$. 

This result shows that the many-particle absorption process of Penrose particles, which are generated inside the ergosphere, cannot be continued arbitrarily, but it is limited by the black hole condition. Moreover, only particles with negative angular momentum are allowed, i.e., particles that rotate in a direction opposite to that of the black hole.

\begin{figure}
    \centering
    \includegraphics[scale=0.4]{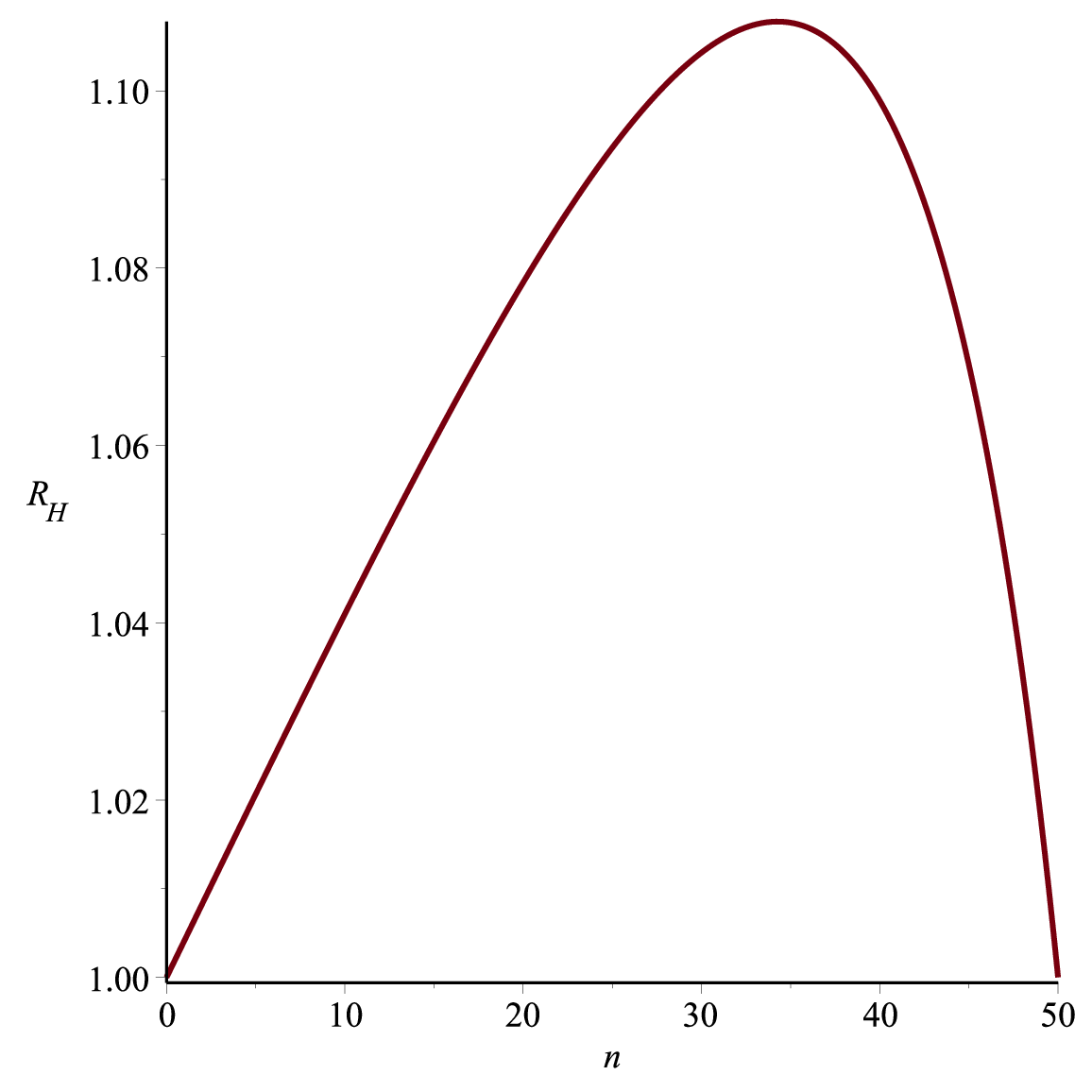}
    \caption{The horizon radius of a rotating black hole after the absorption of $n$ test particles with mass $m=-0.01M$ and angular momentum $j=-0.02M^2$.}
    \label{fig6}
\end{figure}

We will now investigate how the event horizon of an extreme Kerr black hole is affected by the many-particle process, in which each particle follows the dynamics of the Penrose process. According to Eq.(\ref{rhp}), after the absorption of $n$ particles, the the outer horizon radius becomes 
\be
R_H = M+ n m + \frac{\sqrt{(M+nm)^4-(M^2+nj)^2}}{M+nm} .
\ee
In Fig. \ref{fig6}, we show the value of the horizon radius resulting from the absorption of $n$ Penrose particles with negative mass and angular momentum. It can be seen that the horizon radius first increases as particles are absorbed until it reaches a maximum value. Then, it decreases until it reaches the initial state of an extreme Kerr black hole. This second part seems to be nonphysical because the absorbed particles with negative angular momentum tend to diminish the spin of the black hole. In fact, it can be shown that before the horizon radius reaches its maximum value, the absorption process stops because the decay of test particles inside the ergosphere is no longer possible. This has been shown recently in \cite{ruffini2024single} and \cite{ruffini2024role} by analyzing in detail the dynamics of the Penrose process. 

As mentioned before, particles with negative mass can be generated only inside the ergosphere as a result of the decay of test particles. The location of the decay is very important because it determines the value of $m$. In general, the decay can occur only if the condition 
\be
R_E < R_{decay} < R_H 
\ee
is satisfied. Since the absorption process leads first to an increase of $R_H$, after a certain number of absorptions it reaches the location where the decay occurs, $R_H = R_{decay}$, implying that the decay is physically no more realizable. 

Moreover, the increase of the horizon radius is accompanied by a decrease in the ergosphere radius, which after $n$ absorptions is given by 
\be
R_{E,f} = 2 (M_i+nm) .
\ee
Since $m$ is negative, the radius of the ergosphere decreases after each absorption, and, in principle, it can reach the value of $R_{decay}$, making the decay process impossible. Given a value of $R_{decay}$, one can easily calculate the maximum number of allowed absorptions by solving the equations $R_{E,f}=R_{decay}$ or $R_{H,f}=R_{decay}$. The optimal location for the decay is determined by the condition 
\be 
R_{E,f} = R_{decay}=R_{H,f}, 
\ee
which can be solved and yields 
\be
n=-\frac{J_i}{j}.
\ee
Since $j$ is negative, this is always a positive number with $n>>1$ due to the smallness of $j$.

We conclude that in the case of a many-particle absorption process, the horizon radius is an increasing function of the number of absorptions and that there is a maximum number of absorptions determined by the value of the decay radius.

\section{Conclusions}
\label{sec:con}

This work was devoted to the investigation of the absorption of test particles by black holes. The main goal was to understand how the absorption of particles affects the properties of the event horizon. For simplicity, we focus on particles falling radially on the equatorial plane of the black hole. For our analysis, we also take into account the condition under which the falling particles do not lead to the formation of naked singularities. 

We consider particles that arrive from infinity with positive energy, enter the ergosphere, and then cross the event horizon. For this type of particles, we established that the horizon radius always increases as a result of the absorption of particles. 

We also  investigated the absorption process for Penrose particles, which possess negative rest energy as the result of the decay of other particles inside the ergosphere, following the conditions and dynamics of the original Penrose process. 

Moreover, we considered the many-particle absorption process, in which at each step a particle is absorbed, guaranteeing the test particle approximation. We found that the black hole condition implies that there is maximum number of absorptions, which depends on the parameters of the incoming particles. 

In the case of particles arriving from infinity with positive rest energy, we found that, as long as the black hole condition is satisfied, the horizon radius always increases. For Penrose particles with negative energy, we established that the horizon radius increases after each absorption until it reaches a maximum value. 
However, if we consider the dynamics of the Penrose process, as described in references \cite{ruffini2024single} and \cite{ruffini2024role}, that maximum value cannot be reached because the horizon radius end up at the location where the decay process occurs, preventing the decay and formation of further particles with negative energy.

We conclude that the absorption of test particles results, in general, in an increase in the value of the horizon radius. Accordingly, other quantities that characterize black holes, such as 
the irreducible mass, defined by the relationship
\cite{1970PhRvL..25.1596C,1971PhRvD...4.3552C}
\be 
M^2 = M_{irr}^2 + \frac{J^2}{4 M_{irr}^2}, 
\label{mirr}
\ee
i.e.,
\be 
M_{irr} ^2 = \frac{M}{2} \left(M + \frac{1}{M}\sqrt{M^4 - J^2}\right) = \frac{1}{2}MR_H 
\ee 
or the entropy defined by \cite{bekenstein1973black,hawking1975particle}
\be 
S=2\pi(M^2+\sqrt{M^4-J^2})= 2\pi MR_H, 
\ee 
are also quantities that increase in value during the absorption of test particles because they depend linearly on the horizon radius.


\section*{Acknowledgments}

I thank the ICRANet Center at Pescara, Italy, for its hospitality during the elaboration of the main part of this research.
This work was supported by DGAPA-PASPA-UNAM and DGAPA-UNAM, grant No. IN108225.

\bibliographystyle{ws-procs961x669}
\bibliography{horizon}
\end{document}